\documentclass[11pt,a4paper]{article}
\usepackage{graphicx}

\begin{document}

\title{Stable Optimization of a Tensor Product Variational State}

\author{Andrej \textsc{Gendiar},$^{1,2}$ Nobuya \textsc{Maeshima},$^3$ and
Tomotoshi \textsc{Nishino}$^1$\\
$^1$Department of Physics, Faculty of Science, Kobe
University,\\ Kobe 657-8501, Japan\\
$^2$Institute of Electrical Engineering, Slovak Academy of Sciences,\\
D\'ubravsk\'a cesta 9, SK-842 39 Bratislava, Slovakia\\
$^3$Department of Physics, Graduate School of Science, Osaka University,\\
Toyonaka 560-0043, Japan}

\maketitle

\begin{abstract}
We consider a variational problem for three-dimensional (3D)
classical lattice models. We construct the trial state as a 
two-dimensional product of local variational weights that contain
auxiliary variables. We propose a stable numerical algorithm for
the maximization of the variational partition function per layer.
The numerical stability and efficiency of the new method are
examined through its application to the 3D Ising model.
\end{abstract}

\section{Introduction}

The density matrix renormalization group (DMRG) has been applied to a
variety of one-dimensional (1D) quantum systems and 2D classical
systems~\cite{White,Nishino,DMRG}. This method has also been applied to
finite-size 2D quantum systems, which can be represented as 1D quantum
systems with long-range interactions.~\cite{Liang,DMRG} Despite this
success, no extension of the DMRG to {\it infinitely large}
higher-dimensional systems has been reported to this time. This is partially
because the density matrix eigenvalues decay very slowly in higher-dimensional
systems, and the RG transformation, {\it which is obtained from the
diagonalization of the density matrix}, becomes ineffective.~\cite{Peschel}

The numerical efficiency of the DMRG for 1D quantum and 2D classical systems
is due in part to its variational background,~\cite{Ost,Rom,Sierra} in which
the trial state is constructed as a product of orthogonal matrices that 
represent the block-spin (or the renormalization group) transformations.
Such a construction of the variational (or trial) state can be
generalized to higher dimensions. A simple example is to use a 2D classical
system as a variational state for 2D quantum and 3D classical systems. 
Mart\'{\i}n-Delgado {\it et al.} employed the 6-vertex model as a trial wave function 
for 2D lattice spin/electron systems.~\cite{Sierra_st} Okunishi and Nishino
considered the direct extension of the Kramers-Wannier approximation~\cite{KWA}
to the 3D Ising model, in which the variational state is the 2D Ising model
subject to an effective magnetic field.~\cite{KWON} These examples have
demonstrated the potential usefulness of the 2D variational state, constructed
as a product of local variational weights.

For the purpose of obtaining a better variational result, Nishino {\it et al.}
developed a numerical method, the tensor product variational approach 
(TPVA), which automatically optimizes the 2D variational state using
the solution of a self-consistent equation.~\cite{TPVA1} They applied 
this method to both the 3-state ($q = 3$) Potts and the Ising models on
simple cubic lattices, and reported that TPVA yields a better estimate of the
transition temperature for the $q = 3$ Potts model. This is because the
number of variational parameters in the trial state for the Potts model,
$3^4_{~}$, is greater than that for the Ising model, $2^4_{~}$.

In general, the TPVA gives a lower variational free energy when the trial
state contains a larger number of parameters. A way of increasing this number is 
to employ a variational state that contains auxiliary variables, for example
block spin variables.~\cite{TPVA2}\footnote{There is another way of tuning
the local variational weight by way of the density matrix renormalization
in vertical direction (see Ref.~\cite{TPS}).} This generalization, however, 
introduces a serious instability into the numerical optimization of the
variational state.~\cite{TPVA2}

In this paper we report on a method of stabilizing the numerical optimization
process. We introduce an orthogonal condition placed on a small change of
the local variational weight. This condition prevents an `unexpected decrease'
of the norm of the variational state, which caused numerical instability in
previous calculations.~\cite{TPVA2} In the next section, we review the
formulation of the TPVA.~\cite{TPVA2,TPVA1,TPVA3} In \S 3 we consider the
stability of the optimization process. The numerical efficiency of the
stabilized TPVA is examined in \S 4 in the case that it is applied to the
3D Ising model. Conclusions are given in \S 5.

\section{Tensor product variational approach}

\begin{figure}[tb]
\centerline{\includegraphics[width=85mm,clip]{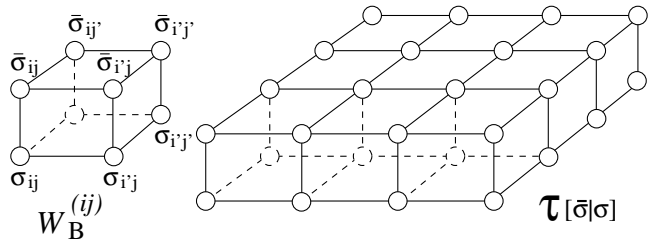}}
\caption{The Boltzmann weight
$W_{\rm B}^{(ij)}\{ \bar\sigma | \sigma \}$ and the transfer matrix
${\cal T}[ \bar\sigma | \sigma ]$ in the case $N=2$.}
\label{tm}
\end{figure}

We consider the 3D Ising model on a simple cubic lattice as an
example. The system size is $2N \times 2N \times \infty$ in the
$X$-, $Y$-, and $Z$-directions. On each lattice point there is
an Ising spin $\sigma$, with $\sigma = \pm 1$. A ferromagnetic
interaction $-J \sigma \sigma'$ exists between nearest-neighbor
spins.

As shown in Fig.~\ref{tm}, the transfer matrix ${\cal T}$
connects adjacent spin layers $[ \sigma ]$ and $[ \bar\sigma ]$,
where $[ \sigma ]$ represents all the spins in a layer:
\begin{equation}
[ \sigma ] = \{ \sigma_{1\,1}^{~}, \sigma_{2\,1}^{~}, \cdots,
\sigma_{2N\,1}^{~}, \sigma_{1\,2}^{~}, \sigma_{2\,2}^{~}, \cdots,
\sigma_{2N\,2N}^{~} \} \,.
\label{sm}
\end{equation}
For simplicity, we consider a symmetric transfer matrix,
\begin{equation}
{\cal T}[ \bar\sigma | \sigma ] = \prod_{i,j = 1}^{2N - 1}
W_{\rm B}^{(ij)}\{ \bar\sigma | \sigma \} \, ,
\label{irfeqtm}
\end{equation}
where $W_{\rm B}^{(ij)}\{ \bar\sigma | \sigma \}$ represents a local
Boltzmann weight with respect to a unit cube at the position
$(X,Y)=(i,j)$ (see Fig.~\ref{tm}). We denote a spin plaquett,
a group of the nearest 4 spins, as
\begin{equation}
\{ \sigma \}
= (\sigma_{i j}^{~}\ \sigma_{i j+1}^{~}
\ \sigma_{i+1 j}^{~}\ \sigma_{i+1 j+1}^{~})
= (\sigma_{ij}^{~}\ \sigma_{ij^\prime}^{~}
\ \sigma_{i^\prime j}^{~}\ \sigma_{i^\prime j^\prime}^{~}),
\end{equation}
using the $i^\prime = i + 1$ and $j^\prime = j + 1$.
With this notation, we can write the local Boltzmann weight
of the 3D Ising model as follows:
\begin{eqnarray}
\nonumber
W_{\rm B}^{(ij)}\{ \bar\sigma | \sigma \}
& = & \exp \biggl[-\frac{J}{4 k_{\rm B}^{~} T} \left(
\sigma_{ij}^{~}\sigma_{i'j}^{~}+\sigma_{i'j}^{~}\sigma_{i'j'}^{~}
+\sigma_{i'j'}^{~}\sigma_{ij'}^{~}
+\sigma_{ij'}^{~}\sigma_{ij}^{~}+\bar\sigma_{ij}^{~}\bar\sigma_{i'j}^{~}
\right. \\
\label{WB3D}
\nonumber
& + &
\bar\sigma_{i'j}^{~}\bar\sigma_{i'j'}^{~}
+\bar\sigma_{i'j'}^{~}\bar\sigma_{ij'}^{~}+
\bar\sigma_{ij'}^{~}\bar\sigma_{ij}^{~}\left.
+\sigma_{ij}^{~}\bar\sigma_{ij}^{~}+
\sigma_{i'j}^{~}\bar\sigma_{i'j}^{~}+
\sigma_{i'j'}^{~}\bar\sigma_{i'j'}^{~}+
\sigma_{ij'}^{~}\bar\sigma_{ij'}^{~}
\right)\biggr] \, .
\end{eqnarray}
We have thus expressed the 3D Ising model as a special case of
the `interaction round a face' (IRF) model.

\begin{figure}[tb]
\centerline{\includegraphics[width=85mm,clip]{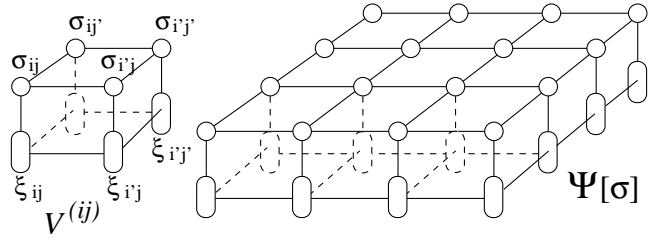}}
\caption{A graphical representation of the local tensor $V^{(ij)}_{~}$
and the construction of the trial state $\Psi$ in the case $N = 2$.
The circles and ovals denote the Ising spins $\sigma$ and the $n$-state
auxiliary variables $\xi$, respectively.}
\label{vs}
\end{figure}

For an arbitrary variational function $\Psi$, the Rayleigh ratio
\begin{equation}
\lambda( \Psi ) = \frac{\sum\limits_{[ \bar\sigma ], [ \sigma ]}
\Psi\left[ \bar\sigma \right] {\cal T}[ \bar\sigma | \sigma ]
\Psi\left[ \sigma\right ]}{\sum\limits_{[ \bar\sigma ], [ \sigma ]}
\Psi\left[ \bar\sigma\right ] \Psi\left[ \sigma\right ]}
\equiv\frac{\langle \Psi | {\cal T} | \Psi \rangle}{\langle \Psi |
\Psi \rangle}
\label{rr}
\end{equation}
gives the variational partition function per layer. In the framework
of the TPVA, the trial state $\Psi$ is constructed as a contracted product
of local variational weights as
\begin{equation}
\Psi\left[ \sigma\right ]
\, = \, \sum\limits_{[\xi]}^{~}\prod\limits_{i,j=1}^{2N-1}
V\left( {{\sigma_{ij}^{~}}\atop{\xi_{ij}^{~}}}\ {{\sigma_{i^\prime
j}^{~}}\atop{\xi_{i^\prime j}^{~}}}
\ {{\sigma_{ij^\prime}^{~}}\atop{\xi_{ij^\prime}^{~}}}\
{{\sigma_{i^\prime j^\prime}^{~}}
\atop{\xi_{i^\prime j^\prime}^{~}}} \right)
\, \equiv \, \sum\limits_{[ \xi ]}
\prod\limits_{i,j=1}^{2N-1}V^{(ij)}\left( {{\{ \sigma \}}\atop{\{ \xi
\}}} \right) \, ,
\label{vsIRF}
\end{equation}
where the variables $\xi_{ij}^{~}$ of the local weight $V$ are auxiliary
variables that can be in one of $n$ states ($0,1,\ldots,n-1$), while the
spin variables $\sigma_{ij}^{~}$ can be in one of 2 states
($\pm 1$)~\footnote{Two kinds of tensor products are known. One
is the IRF type~\cite{TPVA1} and the other is the vertex type.~\cite{TPVA2}
We use the former throughout this section.} (see Fig.~\ref{vs}). We have
expressed a group of auxiliary variables as $[ \xi ]$, analogously to
$[ \sigma ]$ in Eq.~(\ref{sm}).~\footnote{When $n = 1$, the variational
function $\Psi$ does not contain any auxiliary variables.} We are interested
in the bulk properties of lattice models, and therefore we consider
the case in which the system size $2N$ is sufficiently large and
assume that the local variational weight $V^{(ij)}_{~}$ is position
independent. Hereafter, we refer to this variational state as the 
`tensor product state' (TPS).

Because both the trial state $\Psi$ and the transfer matrix ${\cal T}$ are
written as products of local variational weights, both the numerator 
and denominator of Eq.~(\ref{rr}) can also be expressed as products
of stacked local weights:
\begin{eqnarray}
\langle \Psi | {\cal T} | \Psi \rangle & = & \sum\limits_{{[
\bar\sigma ],[ \sigma ]}
\atop{[ \bar\xi ], [ \xi ]}}\prod\limits_{i,j=1}^{2N-1}
V^{(ij)}_{~}\left( {{\{ \bar\sigma \}}\atop{\{ \bar\xi \}}} \right)
W_{\rm B}^{(ij)}\{ \bar\sigma | \sigma \}\
V^{(ij)}_{~}\left( {{\{ \sigma \}}\atop{\{ \xi \}}} \right) \, , \nonumber\\
\langle \Psi | \Psi \rangle & = & \sum\limits_{{[ \sigma ]}
\atop{[ \bar\xi ], [ \xi ]}}\prod\limits_{i,j=1}^{2N-1}
V^{(ij)}_{~}\left( {{\{ \sigma \}}\atop{\{ \bar\xi \}}} \right)\
V^{(ij)}_{~}\left( {{\{ \sigma \}}\atop{\{ \xi \}}} \right) \, .
\label{vv}
\end{eqnarray}
In other words, both $\langle \Psi | \Psi \rangle$ and $\langle \Psi |
{\cal T} | \Psi \rangle$ are partition functions of stacked two-dimensional
classical systems.
This structure enables us to calculate $\langle \Psi | \Psi \rangle$ and
$\langle \Psi | {\cal T} | \Psi \rangle$ accurately by use of numerical
renormalization techniques.~\cite{CTMRG1,CTMRG2,Nishino} Thus, the
variational partition function $\lambda( \Psi )$ can be easily calculated
for an arbitrary TPS constructed from the local variational weight $V$.

\begin{figure}[tb]
\centerline{\includegraphics[width=130mm,clip]{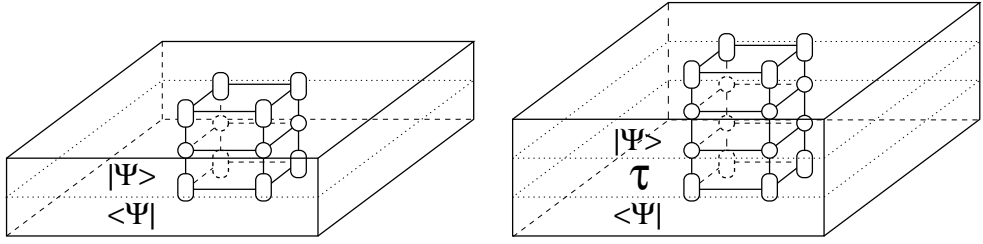}}
\caption{Graphical representation of the `matrices' ${\cal A}$
(on the left) and ${\cal B}$ (on the right) in Eqs.~(\ref{eqa}) and
(\ref{eqb}), respectively.}
\label{ab}
\end{figure}

We maximize $\lambda( \Psi )$ by tuning elements of the local weight $V$.
In order to clarify the variational structure with respect to $V$, let
us divide $\langle \Psi | \Psi \rangle$ into two parts: ({\bf i}) the
adjacent local weights $V^{(NN)}_{~}$ and $\bar V^{(NN)}_{~}$ at the center,
and ({\bf ii}) the rest, consisting of the stacked classical system with
a puncture at the center~\cite{Martin}, expressed as
\begin{equation}
{\cal A}\left( {{\{ \sigma \}}\atop{\{ \bar\xi \}}}\left | {{\{
\sigma \}}\atop{\{ \xi \}}}
\right. \right) = \sum\limits_{{[ \sigma ]^\prime}\atop{[ \bar\xi
]^\prime, [ \xi ]^\prime}}
\prod\limits_{(i,j)\neq(N,N)}V^{(ij)}_{~}\left( {{\{ \sigma \}}
\atop{\{ \bar\xi \}}} \right)V^{(ij)}_{~}\left( {{\{ \sigma \}}\atop{\{
\xi \}}} \right) \, ,
\label{eqa}
\end{equation}
where the configuration sums are taken over all values of the variables
$\sigma$ and $\xi$, except for those at the center that are variables
of $V^{(NN)}_{~}$ and $\bar V^{(NN)}_{~}$.
The notation $\prod_{(i,j)\neq(N,N)}$ indicates that $V^{(NN)}_{~}$
is not included in the product. In the same manner, we divide
$\langle \Psi | {\cal T} |\Psi \rangle$ into ({\bf i'}) $V^{(NN)}_{~}$
and $\bar V^{(NN)}_{~}$, and ({\bf iii}) the rest, expressed by
\begin{eqnarray}
\nonumber
{\cal B}\left({{\{ \bar\sigma \}}\atop{\{ \bar\xi \}}}\left |
{{\{ \sigma \}}\atop{\{ \xi \}}}\right.\right)
& = &
W_{\rm B}^{(NN)}\{ \bar\sigma | \sigma \}\sum\limits_{{[ \bar\sigma ]^\prime,
[ \sigma ]^\prime}\atop{[ \bar\xi ]^\prime, [ \xi
]^\prime}}\prod\limits_{(i,j)\neq(N,N)}
V^{(ij)}_{~}\left({{\{ \bar\sigma \}}\atop{\{ \bar\xi \}}}\right)\\
&\times&W_{\rm B}^{(ij)}\{ \bar\sigma | \sigma \}V^{(ij)}_{~}\left({{\{ \sigma \}}
\atop{\{ \xi \}}}\right),
\label{eqb}
\end{eqnarray}
which corresponds to a partially punctured stacked classical system.
The meanings of the matrices ${\cal A}$ and ${\cal B}$ are represented
graphically in Fig.~\ref{ab}. Using this new notation, we can rewrite
the Rayleigh ratio, Eq.~(\ref{rr}), as the following ratio of bilinear
forms:
\begin{equation}
\label{vbvvav}
\lambda( \Psi )
  \, = \,
\frac{\sum\limits_{{\{ \bar\sigma \}, \{ \sigma \}}\atop{\{ \bar\xi
\},\{ \xi \}}}
V^{(NN)}_{~}\left({{\{ \bar\sigma \}}\atop{\{ \bar\xi \}}}\right)
{\cal B}\left({{\{ \bar\sigma \}}\atop{\{ \bar\xi \}}}
\left | {{\{ \sigma \}}\atop{\{ \xi
\}}}\right.\right)V^{(NN)}_{~}\left({{\{ \sigma \}}
\atop{\{ \xi \}}}\right)}
{\sum\limits_{{\{ \sigma \}}\atop{\{ \bar\xi \},\{ \xi \}}}
V^{(NN)}_{~}\left({{\{ \sigma \}}\atop{\{ \bar\xi \}}}\right)
{\cal A}\left({{\{ \sigma \}}\atop{\{ \bar\xi \}}}
\left|{{\{ \sigma \}}\atop{\{ \xi
\}}}\right.\right)V^{(NN)}_{~}\left({{\{ \sigma \}}
\atop{\{ \xi \}}}\right)}
\, \equiv \, \frac{( V | {\cal B} | V )}{( V | {\cal A} | V )} \, .
\end{equation}
Here, we have interpreted the variational weight at the center
$V^{(NN)}_{~}$ as a $(2n)^4_{~}$-dimensional vector and have written
it simply as $| V )$.

As we have assumed that the system size $2N$ is sufficiently large,
the variation of $\lambda( \Psi )$ with respect to a uniform
modification of local weights,
\begin{equation}
\delta\lambda( \Psi ) = \sum_{i j}^{~}
\frac{\partial \lambda( \psi )}{\partial V^{(ij)}_{~}}
\delta V^{(ij)}_{~}+ {\rm [higher~order~terms]},
\end{equation}
is almost proportional to $\partial \lambda( \psi )
/ \partial V^{(NN)}_{~}$, which is the contribution from the
local variation only at the  center.~\cite{TPVA1,TPVA2}
From the optimal condition
$\partial\lambda( \psi ) / \partial V^{(NN)}_{~} = 0$, we obtain
the generalized eigenvalue problem
\begin{equation}
{\cal B} \, | V ) \, = \, \lambda {\cal A} \, | V ) \, .
\label{sce}
\end{equation}
We use this equation when we optimize the TPS. Note that
Eq.~(\ref{sce}) is a non-linear equation with respect to the
local weights $V$, since the $(2n)^4_{~}$-dimensional matrices
${\cal A}$ and ${\cal B}$ themselves are functionals of $V^{(NN)}_{~}$.
Therefore, Eq.~(\ref{sce}) should be solved self-consistently
by use of successive, small improvements of the local variational
weight $V^{(NN)}_{~}$.

In Ref.~\cite{TPVA2}, Nishino {\it et al.} apply the TPVA to the
3D Ising model, which is represented as a symmetric 16-vertex model.
For the case $n = 2$, they succeeded in optimizing the TPS.
However, they encountered numerical instability in the case
$n \geq 3$. The reason for this is that the matrix ${\cal A}$ is
not always positive definite during the process in which the variational
state is being improved, although ${\cal A}$ should be positive definite
after the optimization is completed. We discuss the cause of this
instability and propose a method of stabilization in the following.

\section{Stabilization of the self-consistent improvement process}

Consider an infinitesimal change of the local weight at the center
of the system,
\begin{equation}
| V ) \, \rightarrow \, | V' ) \, = \, | V ) \, + \, \varepsilon | X ),
\label{VX}
\end{equation}
where $| X )$ is an arbitrary $(2n)^4_{~}$-dimensional vector. The Rayleigh
ratio, Eq.~(\ref{vbvvav}), is modified as
\begin{equation}
\lambda' \, \equiv \, \frac{( V' | {\cal B} | V' )}{( V' | {\cal A} | V' )}
               \, = \, \frac{( V  | {\cal B} | V  )}{( V  | {\cal A} | V  )}
\left(
\frac{\displaystyle 1 + 2 \varepsilon\frac{( V | {\cal B} | X )}{( V |
{\cal B} | V )}
+ \varepsilon^2\frac{( X | {\cal B} | X )}{( V | {\cal B} | V )}}
{\displaystyle 1 + 2\varepsilon \frac{( V |{\cal A} | X )}{( V |
{\cal A} | V )}
+ \varepsilon^2\frac{( X | {\cal A} | X )}{( V | {\cal A} | V )}}\right) \, .
\label{der1}
\end{equation}
If $\varepsilon$ is sufficiently small, we can expand $\lambda'$ as
\begin{equation}
\lambda' = \lambda \left[ 1 + 2 \varepsilon ( \beta - \alpha )
  + {\cal O}(\varepsilon^2)\right],
\label{der2}
\end{equation}
with
\begin{equation}
\lambda = \frac{( V | {\cal B} | V )}{( V | {\cal A} | V )} \, ,\quad
\alpha  = \frac{( V | {\cal A} | X )}{( V | {\cal A} | V )} \, ,\quad
\beta   = \frac{( V | {\cal B} | X )}{( V | {\cal B} | V )} \, .
\end{equation}
It seems appropriate to select $| X )$ to maximize $\beta - \alpha$.
However, such a choice of $| X )$ tends to cause the expectation
value of the denominator ${( V' | {\cal A} | V' )}$ to decrease,
and after several self-consistent iterations, ${( V | {\cal A} | V )}$
becomes very small or negative. This is a cause of numerical instability
in the previous TPVA algorithm. We prevent this {\it anomalous decrease}
of $( V | {\cal A} | V )$ by choosing $| X )$ to satisfy
\begin{equation}
( V' | {\cal A} | V' ) \, = \, ( V | {\cal A} | V )
                           + {\cal O}(\varepsilon^2_{~}) \, .
\label{shrink}
\end{equation}
This is equivalent to choosing $| X )$ to be orthogonal to
${\cal A} | V )$, which gives $\alpha = 0$. With such a choice,
the maximization of $\lambda$ is carried out with respect to
$\beta$ only. A reasonable choice of $| X )$ can be obtained using
the Schmidt orthogonalization:
\begin{equation}
| X ) = {\cal B} | V ) - {\cal A} | V ) \, ( V | {\cal AB} | V ) \, .
\label{orthog}
\end{equation}
The $| X )$ thus obtained at least yields non-negative $\beta$. We
include $| X )$ in the self-consistent calculation of the TPVA as follows:
\begin{itemize}
\item[(1)] Prepare an arbitrary initial variational weight $| V )$.
\item[(2)] Calculate the matrices ${\cal A}$ and ${\cal B}$ using
a numerical RG method.
\item[(3)] Obtain $| X )$ from Eq.~(\ref{orthog}).
\item[(4)] Improve the local weight according to $| V^{\prime} ) = | V ) +
\varepsilon | X )$.
\item[(5)] Terminate the modification of $| V )$ if the computed
thermodynamic functions have converged; otherwise go back to step (2).
\end{itemize}
The third step is the main difference between the present and the previous
TPVA algorithms. Since $( V | {\cal A} | V )$ changes only by an amount of
order $\varepsilon^2_{~}$ at each iteration, a large number of iterations
is necessary to realize convergence, especially when $\varepsilon$ is small.

\section{Numerical results}

We now confirm the numerical stability of the new algorithm in the case that
it is applied to the 3D Ising model. Hereafter, we set $J/k_{\rm B}^{~}=-1$
and denote by $m$ the number of the states of the block-spin variable used
in the numerical RG calculation~\cite{CTMRG1,CTMRG2}, which is used for the
calculation of ${\cal A}$ and ${\cal B}$.

\begin{figure}[tb]
\centerline{\includegraphics[width=100mm,clip]{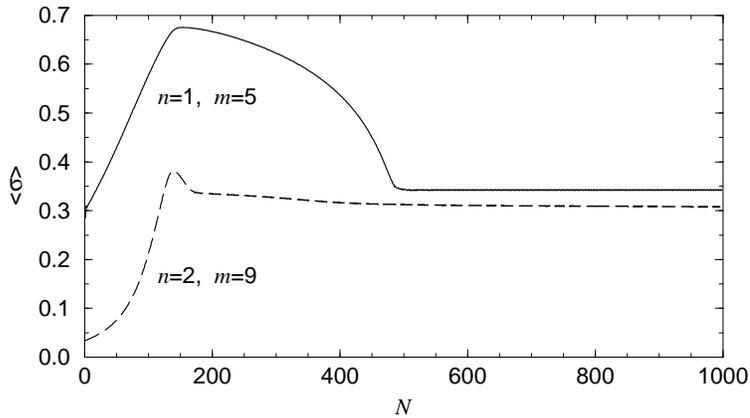}}
\caption{Convergence of the spontaneous magnetization $\langle \sigma \rangle$
with respect to the number of numerical steps at $T = 4.504$.}
\label{Convg}
\end{figure}

Figure~\ref{Convg} shows the convergence of the spontaneous magnetization
$\langle \sigma \rangle$ with respect to the number of numerical iterations
at the temperature $T = 4.504$. (The selected temperature $T=4.504$ is
slightly lower than the critical temperature $T_{\rm c}^{~} = 4.5115$
obtained by Monte Carlo simulations~\cite{Mc} which is considered to be
the most reliable result.) The parameter $\varepsilon$ in
Eq.~(\ref{VX}) is chosen to be $10^{-3}_{~}$ for normalized $| V )$ and
$| X )$. In both the cases $n = 1$ and $n = 2$, $\langle \sigma \rangle$
begins to converge monotonically to the final value after several hundred iterations.

\begin{figure}[tb]
\centerline{\includegraphics[width=67mm,clip]{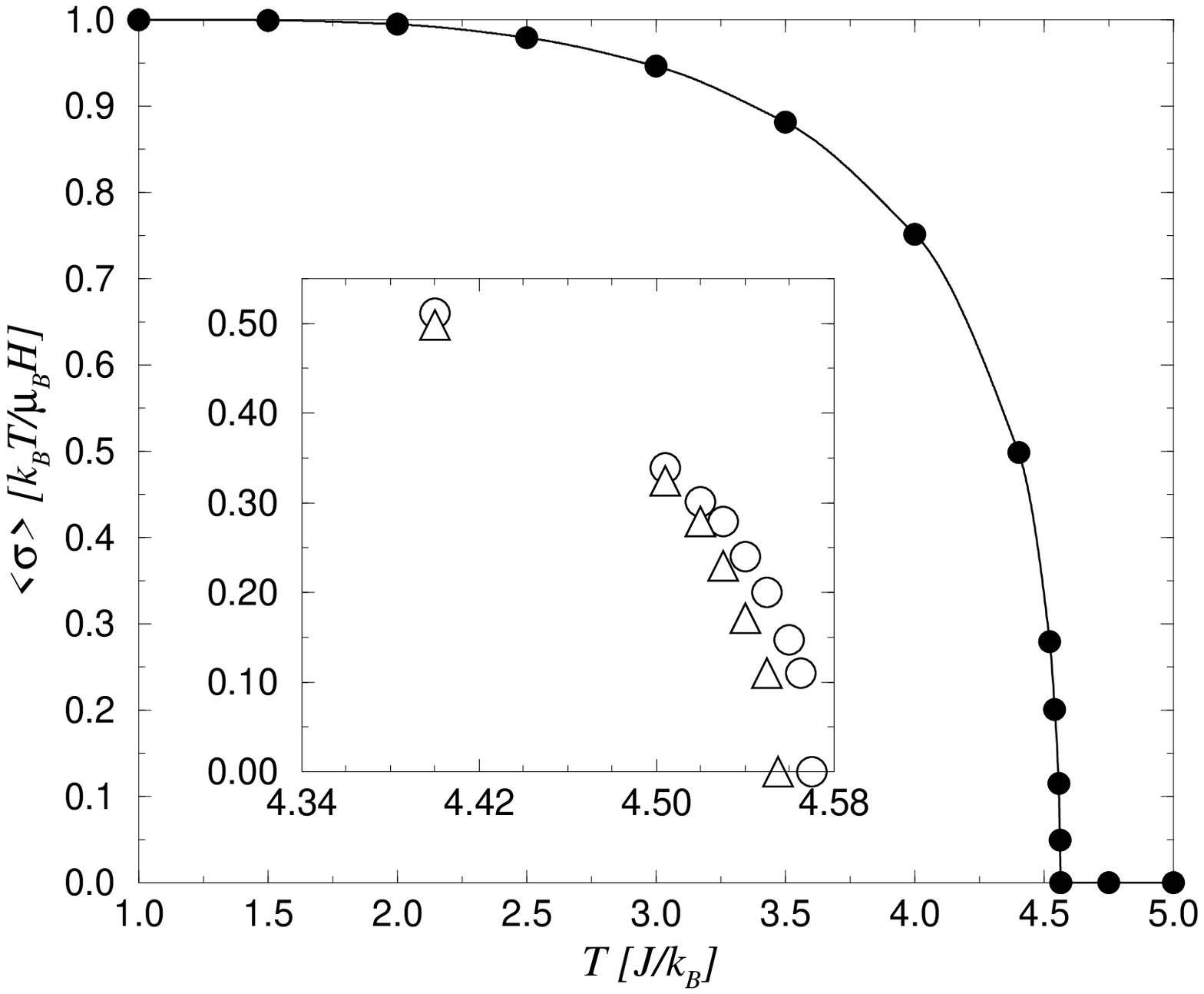}
\includegraphics[width=67mm,clip]{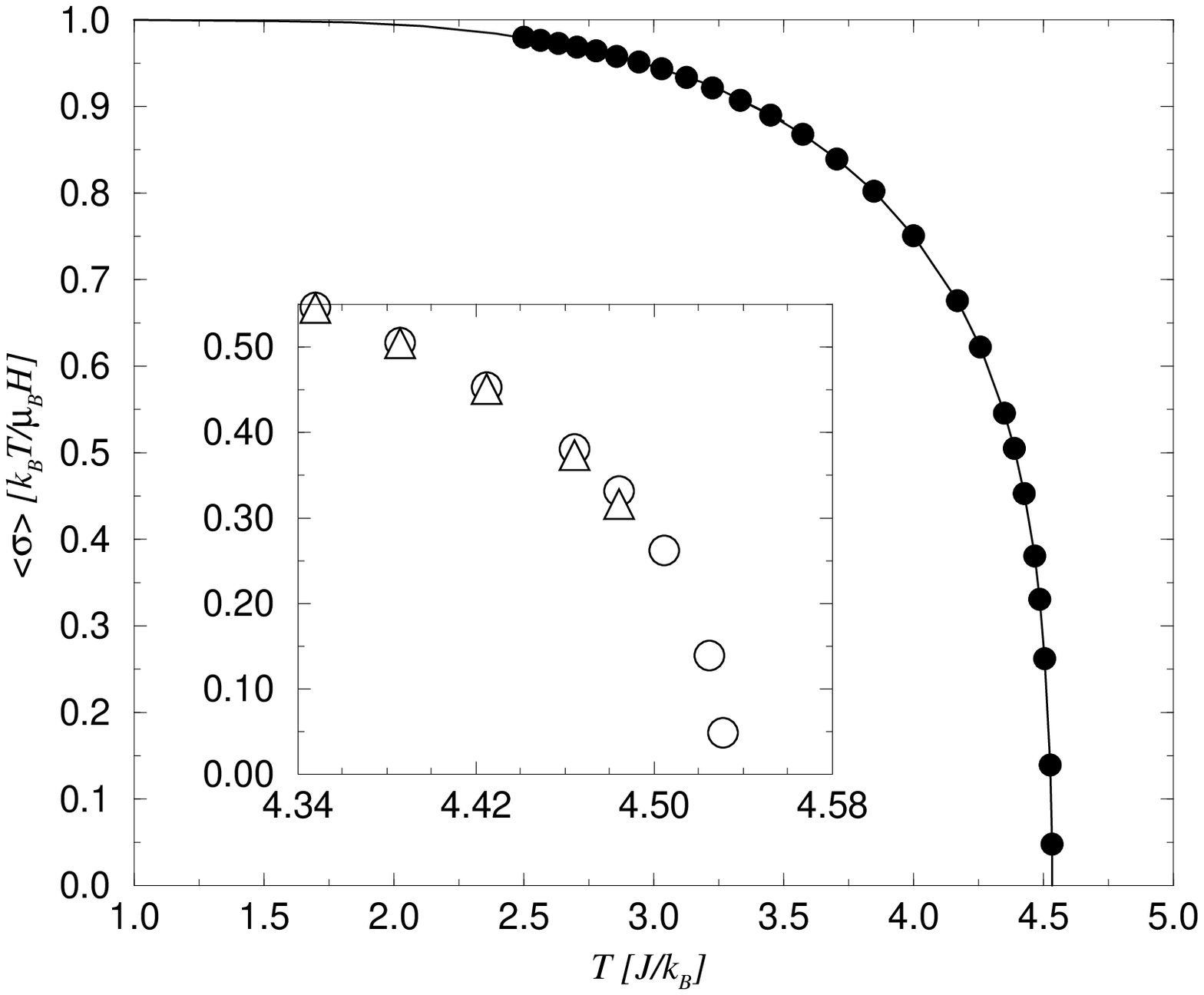}}
\caption{Calculated spontaneous magnetizations. (a) $\langle\sigma\rangle$
obtained for $( n, m ) = ( 2, 5 )$. The curve here was drawn using the Spline
interpolation. The inset displays $\langle\sigma\rangle$ near the criticality
for $( n, m ) = ( 1, 5 )$ (circles) and $( 2, 5 )$ (triangles).
(b) $\langle\sigma\rangle$ obtained from the TPVA applied to the Ising model
represented as the 64-vertex model~\cite{TPVA2} for $( n, m ) = ( 2, 16 )$.
The inset plots the data for $( n, m ) = ( 2, 16 )$ (circles) and
$( n, m ) = ( 3, 12 )$ (triangles).}
\label{IsingIRF}
\end{figure}

Figure~\ref{IsingIRF}(a) displays $\langle\sigma\rangle$ thus calculated
for $n = 1$ (no auxiliary variables) and $n = 2$ with $m = 5$.
In the region $T < 4.2$, the difference between these cases is not
visible. As shown in the inset, near the critical temperature,
the calculated $\langle \sigma \rangle$ with $n = 2$
decays more rapidly than that with $n = 1$. The estimated transition
temperatures, where the obtained $\langle \sigma \rangle$ falls to zero,
are $T = 4.57$ ($n = 1$) and 4.55 ($n = 2$).

To this point, we have expressed the 3D Ising model as a 3D IRF model.
The Ising model can also be expressed as a special case of the symmetric
64-vertex model.~\cite{TPVA2} Applying the stabilized TPVA algorithm to
this vertex expression, we obtain the $\langle \sigma \rangle$ plotted
in Fig.~\ref{IsingIRF}(b). In this case, the estimated transition
temperatures are $T = 4.533$ (with $n = 2$ and $m = 16$) and
4.525 (with $n = 3$ and $m = 12$). All of these calculated transition
temperatures are higher than $T_{\rm c}^{\rm MC} = 4.5115$, obtained
by Monte Carlo simulation.~\cite{Mc} It is thus seen that with the TPVA,
the ordered phase tends to be stabilized.

We finally compare the stabilized numerical algorithm considered here with
the previous one.~\cite{TPVA2} The stabilization provided by the Schmidt
orthogonalization, expressed by Eq.~(\ref{orthog}), enables us to perform
the calculation for those cases $n = 2$ (IRF expression)
and $n = 3$ (Vertex expression) for which the previous algorithm does
not yield a convergent result. With regard to the computational time
required to obtain the convergent numerical result for the cases $n = 1$
(IRF expression) and $n = 2$ (Vertex expression), for which the previous
and the stabilized algorithms give the same numerical result, the
stabilized algorithm is about 10 times slower than the previous algorithm.
In the variational calculation employing the TPVA, the numerator
$( V | {\cal B} | V )$ should be maximized and the denominator
$( V | {\cal A} | V )$ should be minimized through successive
improvements of $V$. Though $( V | {\cal B} | V )$ increases rather
rapidly, $( V | {\cal A} | V )$ decreases slowly, because the
stabilization condition  Eq.~(\ref{shrink}) restricts the size of the
change of $( V | {\cal A} | V )$ to order $\varepsilon^2_{~}$.

\section{Conclusion and discussion}

We have proposed a stabilized partition function maximization process
for numerical calculations employing the TPVA, imposing the orthogonality
condition on the small change made to the local variational weight.
This improvement enables us to obtain the magnetization using a TPS
that contains $n$-state auxiliary variables, in particular, with $n = 2$
for an IRF-type TPS and $n = 3$ for a vertex-type TPS.

The orthogonality condition expressed by Eq.~(\ref{orthog}) stabilizes
the numerical calculation, but it also slows the convergence to the
variational maximum, because it causes the improvement of the denominator
of the variational formulation to be very slow. Contrastingly, the vertical
density matrix approach (VDMA)~\cite{TPS}, which also uses a TPS as its
trial state, exhibits rapid convergence. In the VDMA, the local weight
is improved through a RG transformation obtained from the diagonalization
of the density matrix, and thus both the denominator and the numerator of
the variational ratio are improved equally rapidly. The VDMA, however,
has the shortcoming that it does not fully improve the TPS, because the
RG transformation used in the VDMA is not determined so as to maximize
the variational partition function per a layer. {\it In higher dimensions,
the direct diagonalization of the density matrix does not give the most
appropriate RG transformation.} Our next project is to combine the
advantages of the TPVA and the VDMA.

\section*{Acknowledgements}
We thank K.~Okunishi, Y.~Hieida and Y.~Akutsu for valuable discussions.
This work has been partially supported by a Grant-in-Aid for Scientific
Research from the Ministry of Education, Science, Sports and Culture
(Grant No.~09640462 and No.~11640376) and by the Slovak Grant Agencies,
VEGA No.~2/7201/21 and 2/3118/23. A.~G. is also supported by the Japan
Society for the Promotion of Science (P01192).

\end{document}